\documentclass[
    ,final            % use final for the camera ready runs
  ]
  {aipproc}

\layoutstyle{8x11single}

\newcommand{\be}{\begin{equation}}
\newcommand{\ee}{\end{equation}}
\newcommand{\ba}{\begin{eqnarray}} 
\newcommand{\ea}{\end{eqnarray}}

\begin{document} 

\title{Interacting Dark Matter as an Alternative to Dark Energy}

\classification{98.80.-k, 95.35.+d,95.36.+x}
\keywords      {Cosmology; dark matter; dark energy}

\author{Spyros Basilakos}{
address={Academy of Athens, Research Center 
for Astronomy and Applied Mathematics,
 Soranou Efesiou 4, 11527, Athens, Greece}
}

\author{Manolis Plionis}{
address={Institute of Astronomy \& Astrophysics,
  National Observatory of Athens, Palaia Penteli 152 36, Athens, Greece \\
Instituto Nacional de Astrof\'isica, \'Optica y Electronica, Puebla, Mexico}
}
\begin{abstract}
We investigate the global dynamics of the universe within the framework of the 
Interacting Dark Matter (IDM) scenario.
Considering that the dark matter obeys the collisional Boltzmann equation, we can obtain analytical 
solutions of the global density evolution, which can accommodate an accelerated
expansion, equivalent to either the {\em quintessence} or the 
standard $\Lambda$ models. This is possible if there is a
disequilibrium between the DM particle creation and annihilation processes
with the former process dominating, which creates an effective source
term with negative pressure. Comparing the predicted Hubble expansion
of one of the IDM models (the simplest) with observational data, we find that the effective
annihilation term is quite small, as suggested by various experiments.
\end{abstract}

\maketitle

\section{Introduction}
The detailed analysis of the available high quality cosmological 
observations (\cite{Spergel07,essence,Kowal08,komatsu08} and references therein)
have converged during the last decade towards a cosmic expansion
history that involves a spatial flat geometry and 
a recent accelerating expansion of the
universe. This expansion has been attributed to an energy component
(the so called dark energy) with negative pressure which dominates the universe at
late times and causes the observed accelerating expansion. 
The nature of the dark energy is still a mystery and it is 
one of the most fundamental current problems in physics and cosmology. 
Indeed, due to the absence of a physically well-established fundamental theory, there have 
been many theoretical speculations regarding the nature of the above exotic dark energy (DE)
among which a cosmological constant, scalar or vector fields
(see \cite{Weinberg89,Wetterich:1994bg,
Caldwell98,Peebles03,Brax:1999gp,Brookfield:2005td, Boehmer:2007qa} and 
references therein).

Most of the recent papers in this kind of studies are based on the assumption that
the DE evolves independently of the dark matter (DM). The unknown nature
of both DM and DE implies that we can not preclude future surprises 
regarding the interactions in the dark sector. 
This is very important because interactions between the DM and 
{\em quintessence} could provide possible solutions 
to the cosmological coincidence problem. 
Recently, several papers have been published in this area 
\cite{Amm:1999, Bin:2006}
proposing that the DE and DM could be coupled, assuming also that 
there is only one type of non-interacting DM. 

However, there are other possibilities. (a) It is plausible that the dark
matter is self-interacting (IDM) 
\cite{SperStei}, a possibility that has been proposed to solve discrepancies between 
theoretical predictions and astrophysical observations, 
among which the gamma-ray and microwave emission 
from the center of our galaxy 
(eg. \cite{92, 90}, \cite{hoop, regis} and references therein). 
It has also been shown that some dark matter interactions 
could provide an accelerated expansion phase of the
Universe \cite{zim, balakin, Lima2008}.
(b) The DM could potentially contain more than one particle species,
for example a mixture of cold and warm or hot dark matter 
\cite{Farrar:2003uw,Gubser:2004uh}, with or without inter-component interactions.

In this work we are not concerned with the viability of the different 
such possibilities, nor 
with the properties of interacting DM models.
The aim of this work is to investigate only whether there are repercussions of DM
self-interactions for the global dynamics of the universe and specifically whether
such models can yield 
an accelerated phase of the cosmic expansion, without the need of the {\em dark energy}.
We note that we do not ``design'' the fluid interactions to
produce the desired accelerated cosmic evolution, as in some previous works
(eg., \cite{balakin}), but investigate the circumstances
under which 
the analytical solution space of the collisional Boltzmann equation, in
the expanding universe, allows for a late accelerated phase of the
universe.

\section{Collisional Boltzmann Equation in the Expanding Universe}
As we described in the introduction the DM is usually
considered to contain only one type of particles that 
are stable and neutral. In this work, we would like to investigate the cosmological
potential of a different scenario, in which the dominant ''cosmic'' fluid 
does not contain dark energy, 
is not perfect, and at the same time is not in equilibrium.
In this context, initially the total energy density is $\rho=\rho_{\rm IDM}+\rho_{r}$. 
We consider that the self-interacting dark matter does not interact significantly 
with the background radiation, and thus in the matter-dominated epoch, radiation
is irrelevant to the global dynamics (because of the well-known dependence: 
$\rho_{r} \propto a^{-4}$). 
Therefore, taking the latter considerations into account and assuming that there are no 
residual radiation products of the DM interactions (otherwise see \cite{Basplio09}),
we conclude that 
in the matter-dominated era the total cosmic dark-matter density reduces 
to that of the IDM density ($\rho\simeq \rho_{\rm IDM}$), 
which obeys the collisional Boltzmann
equation \cite{Kolb1990}: 
\begin{equation}
\frac{{\rm d}\rho}{{\rm d}t}+3H(t)\rho+\kappa \rho^{2}-\Psi=0 \;,
      \label{bol1} 
\end{equation}
where $H(t)\equiv \dot{\alpha}/\alpha$ is the Hubble function,
$\Psi$ is the rate of creation of DM particle pairs, and
$\kappa (\ge 0)$ is given by:
\begin{equation}
\kappa=\frac{\langle \sigma u \rangle}{M_x} \;,
\end{equation}
where $\sigma$ is the cross-section for annihilation, $u$ is the mean particle velocity,
and $M_x$ is the mass of the DM particle.
It is obvious that if we impose in the current analysis 
an effective pressure term such as  
\begin{equation}\label{PRES}
P=(\kappa \rho^2-\Psi)/3 H\;,
\end{equation}
then the collisional Boltzmann equation reduces to the usual fluid
equation: $\dot{\rho}+ 3H (\rho + P)=0$.
Now if the DM is collisionless or the collisional annihilation and pair creation
processes are in equilibrium (ie., $P=0$), the corresponding solution of the 
above differential equation is $\rho \propto \alpha^{-3}$, 
as it should (where $\alpha$ is the scale factor of the universe).
In contrast, for the case of a non-perfect fluid (ie., having a disequilibrium
between the annihilation and particle pair creation processes) it becomes 
clear that we can either have a positive or a negative effective pressure term.

The possibility of a negative such effective pressure term,
ie., the case for which the DM particle creation term is larger 
than the annihilation term ($\kappa \rho^{2}-\Psi<0$),  is of
particular interest for its reprecursions on the global dynamics of
the Universe (see appendix A). 
In the framework of a Boltzmann formalism, a negative
pressure could in general be the outcome 
of dark matter self-interactions, as suggested in 
\cite{zim, balakin},
if an ``anti-frictional'' force is self-consistently exerted 
on the particles of the cosmic fluid.
This possible alternative to dark energy has the caveat of its unknown
exact microphysics, which is also however the case for all dark energy models.
Note, that in the literature, there
is a wealth of papers discussing the DM particle creation processes in 
the context of thermodynamics (for example, eg. \cite{Prigo1989}):

We now proceed in an attempt to analytically solve eq. (1). To this
end we change variables from $t$ to $\alpha$ and thus eq.(\ref{bol1}) 
can be written:
\begin{equation}
\frac{{\rm d}\rho}{{\rm d}\alpha}=f(\alpha)\rho^{2}+g(\alpha)\rho+R(\alpha) \;\;
      \label{bol2} 
\end{equation}
where
\begin{equation}
f(\alpha)=-\frac{\kappa}{\alpha H(\alpha)} \;\;\; g(\alpha)=-\frac{3}{\alpha} \;\;\;
R(\alpha)=\frac{\Psi(\alpha)}{\alpha H(\alpha)}\;.      
      \label{fun1} 
\end{equation}
In general it is not an easy task to solve analytically eq. (\ref{bol2}),
which is a Riccati equation, due to the fact that it is a 
non-linear differential equation. However, eq. (\ref{bol2}) could be 
fully solvable if (and only if) a particular solution is known. Indeed,
we find that for some special cases regarding the functional form of the
interactive term, such as $\Psi=\Psi(\alpha,H)$, we can derive 
analytical solutions (see appendix B). 
We have phenomenologically identified two functional forms for which
we can solve the previous differential equation analytically, 
only one of which is of interest since it provides
$\propto a^{-3}$ dependence of the scale factor. This is:
\begin{equation}
\Psi(\alpha)=\alpha H(\alpha)R(\alpha)=C_{1}(m+3)\alpha^{m}
H(\alpha)+\kappa C_{1}^{2}\alpha^{2m} \;.
      \label{int1} 
\end{equation}  
Although, the above functional form was not
motivated by some physical theory, but rather 
phenomenologically because it provides analytical solutions
to the Boltzmann equation, its exact form can be justified {\em a posteriori}
within the framework of IDM (see Appendix C).

The general solution of equation (\ref{bol2}) using eq.(\ref{int1}) 
for the total energy density is:
\begin{equation}
\rho(\alpha)=C_{1}\alpha^{m}+\frac{\alpha^{-3}F(\alpha)}
{\left[C_2-\int_{1}^{\alpha} x^{-3} f(x) F(x)dx\right]} \;,
      \label{sol1} 
\end{equation}  
where the kernel function $F(\alpha)$ has the form
\begin{equation}
F(\alpha)={\rm exp}
\left[-2 \kappa C_{1}\int_{1}^{\alpha} \frac{x^{m-1}}{H(x)}dx \right] \;\;.
      \label{ker2} 
\end{equation}  
We note that $\kappa C_1$ has units of Gyr$^{-1}$, while 
$m$, $C_{1}$, and $C_2$ are the corresponding constants of the problem. 
Obviously, Eq.(\ref{sol1}) can be rewritten as  
\begin{equation}
\rho(\alpha)=\rho_{c}(\alpha)+\rho^{'}(\alpha) \;,
      \label{gg33} 
\end{equation}
where $\rho_{c}=C_{1}\alpha^{m}$ is the density corresponding to the
residual matter creation that
results from a possible 
disequilibrium between the particle creation and annihilation processes,
while $\rho^{'}$ can be viewed as the energy density of the self-interacting dark
matter particles that are dominated by the annihilation processes.
We note that close to the present epoch as well as at late
enough times ($\alpha \gg 1$), 
the $\rho^{'}$ evolves in a similar way to the usual dark matter 
(see also \cite {wein08}).
Finally, if both $\kappa$ and $\Psi$ tend to zero, the 
above cosmological model reduces to the usual Einstein-deSitter model
($\rho \propto a^{-3}$).

We note that, since $\rho^{'}>0$, the constant $C_{2}$ obeys the restriction
\begin{equation}
C_{2}> G(\alpha)=\int_{1}^{\alpha} x^{-3} f(x) F(x)dx \ge 0 \;\;.
      \label{regg33} 
\end{equation}
Evaluating now Eq.(\ref{sol1}) at the present time ($\alpha=1$, $F(\alpha)=1$), we obtain 
the present-time total cosmic density, which is: 
$\rho_0=C_{1}+1/C_{2}\;$, with $C_{1} \ge 0$ and $C_{2}>0$. 

\subsection{Conditions to have an inflection point}
Using the Hubble function:
$H^2(\alpha)=(\dot{\alpha}/\alpha)^2 = 8 \pi G \rho(\alpha)/3 \;,$
we derive the second derivate of the scale factor:
\begin{equation}\label{dda}
\frac{\ddot{\alpha}}{\alpha} = 
-\frac{4 \pi G}{3} \left(-2\rho -\frac{\dot \rho}{H} \right) \;\;.
\end{equation}
Inserting Eqs.(\ref{bol1}) and
(\ref{PRES}) into Eq.(\ref{dda}), we obtain
\begin{equation}\label{d1}
\frac{\ddot{\alpha}}{\alpha} = -\frac{4 \pi G}{3} \left(\rho +\frac{\kappa \rho^{2}-\Psi}{H} \right)= 
-\frac{4 \pi G}{3} \left(\rho +3P\right) \;\;.
\end{equation}
In order to have an inflection point ($\ddot{\alpha}=0$), 
the expression $\rho+3P=0$ 
should contain 
a real root in the interval: $\alpha \in (0,1)$. 
Thus, with the aid of Eq.(\ref{sol1}),
(\ref{PRES}) and (\ref{int1}), we derive the following condition:
\begin{equation}\label{ZZ}
\frac{\alpha^{-3}F(H+2\kappa C_{1}\alpha^{m})}{C_{2}-G}+
\frac{\kappa \alpha^{-6}F^{2}}{(C_{2}-G)^{2}}-
(m+2)C_{1}\alpha^{m}H =0\;,
\end{equation}
from which we obtain that $m>-2$ (where $C_1>0$, $\kappa
\ge 0$, and $C_2-G>0$).
Evidently, if we parametrize the constant $m$ according to
$m=-3(1+w_{\rm IDM})$, we obtain the condition $w_{\rm IDM}<-1/3$,
which implies that the current cosmological model can be viewed as
a viable {\em quintessence} dark-energy look-alike, as far 
as the global dynamics is concerned.
We remind the reader that the same restriction holds
for the conventional dark energy model in which 
$P_{Q}=w \rho_{Q}$ ($w=const<-1/3$).

\begin{figure}
\includegraphics[height=.4\textheight]{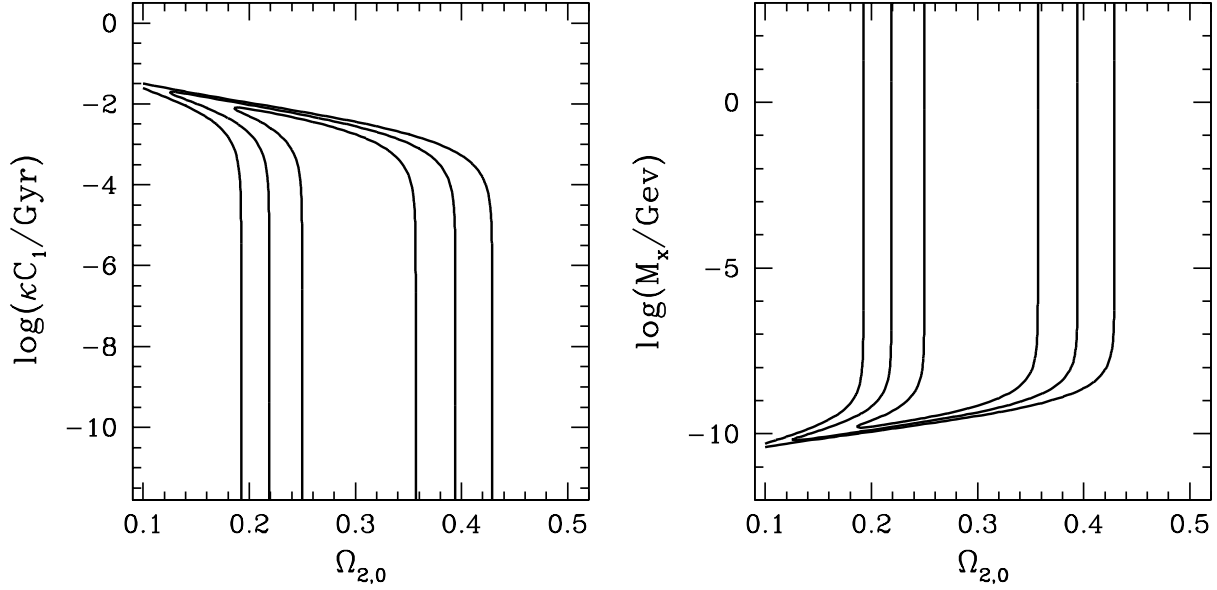}
\caption{{\em Left panel}: The $\Omega_{2,0}-\kappa C_1$ solution
space provided by fitting our model to the early-type galaxy Hubble
relation of Simon et al (2005). {\em Right panel:} The corresponding
$\Omega_{2,0}-M_x$ solution space.}
\end{figure}

\section{Mimicking the $\Lambda$ Cosmology}
We will show that for $m=0$ the global dynamics, provided 
by our model (see eq.\ref{sol1}), is
equivalent to that of the traditional $\Lambda$ cosmology. 
Indeed, we use
${\rm d}t={\rm d}\alpha/ (\alpha H)$ and the basic kernel (Eq. \ref{ker2}) 
becomes
\begin{equation}
F(\alpha)={\rm exp}
\left[-2 \kappa C_{1}\int_{1}^{\alpha} \frac{1}{xH(x)}dx \right]=
e^{-2\kappa C_{1}(t-t_{0})} \;,
      \label{ker4} 
\end{equation}  
where $t_{0}$ is the present age of the universe.
In addition, the integral in 
Eq.(\ref{sol1}) (see also Eq.\ref{regg33}) now takes the form
$G(\alpha)=-\kappa Z(t)$ and 
$Z(t)=\int_{t_{0}}^{t} \alpha^{-3}e^{-2 \kappa C_1 (t-t_0)}$.
We note that at the present time we have $G(1)=0$.
Obviously, using the above formula, 
the global density evolution (Eq. \ref{sol1}) can be written
\begin{equation}
\rho(\alpha)=C_{1}+\alpha^{-3}\frac{e^{-2\kappa C_{1}(t-t_{0})}}
{\left[C_{2}-G(\alpha) \right]} \;\;.
      \label{sol4a} 
\end{equation}  
As expected, at early enough times ($t \rightarrow 0$) the overall density 
scales according to
$\rho(\alpha) \propto a^{-3}$,  while close to the 
present epoch 
the density evolves according to
\begin{equation}\label{eq:lam}
\rho(\alpha)\simeq C_{1}+\frac{\alpha^{-3}}{C_{2}} \;\;,
\end{equation}
which is approximately equivalent to the corresponding evolution in the 
$\Lambda$ cosmology in which the term 
$C_{1}$ resembles the constant-vacuum term ($\rho_{\Lambda}$) 
and the $1/C_{2}$ term resembles the density of matter ($\rho_m$).
We note that the effective pressure term (Eq. \ref{PRES}),
for $\kappa\rightarrow 0$, becomes
$\Psi \sim 3 C_{1} H$, which implies that: $P\sim -\Psi/3 H =  -C_{1}$.
Therefore, this case relates to the traditional 
$\Lambda$ cosmology, since $C_{1}$ corresponds to 
$\rho_{\Lambda}$ (see Eq. \ref{eq:lam}).
We now investigate in detail the dynamics of the $n=0$ 
model.

From Eq.(\ref{sol4a}), using the usual unit-less $\Omega$-like 
parametrization, we derive after some algebra that
\begin{equation}\label{hubb2}
\left(\frac{H}{H_0}\right)^2 = \Omega_{1,0} + \frac{\Omega_{1,0} \Omega_{2,0} \alpha^{-3} 
e^{-2 \kappa C_{1} (t-t_0)} }{\Omega_{1,0} + \kappa C_{1} \Omega_{2,0} Z(t)} \;,
\end{equation}
where $\Omega_{1,0}= 8 \pi G C_{1}/3H_0^2$ and $\Omega_{2,0}=8 \pi G / 3H^2_0 C_{2}$, 
which in the usual $\Lambda$ cosmology relates to $\Omega_{\Lambda}$ and 
$\Omega_{\rm m}$, respectively. 

We can now attempt to compare
the Hubble function of Eq.(\ref{hubb2}) to that corresponding 
to the usual $\Lambda$ model.
To this end, we use a $\chi^2$ minimization between the different
models (our IDM Eq.\ref{hubb2} or the traditional $\Lambda$CDM model) 
and the observed Hubble relation derived 
directly from early-type galaxies at high redshifts 
(\cite {simon05}). 
For the case of our IDM model, we
simultaneously fit the two free
parameters of the model, i.e., $\Omega_{2,0}$ and $\kappa C_{1}$
for a flat background ($\Omega_{1,0}=1-\Omega_{2,0}$) with
$H_{0}=72$ km/sec/Mpc and $t_0=H_0^{-1}\simeq 13.6$ Gyrs which 
is roughly the age of the universe of
the corresponding $\Lambda$ cosmology.
This procedure yields the best-fit model parameters
$\Omega_{2,0}=0.3^{+0.05}_{-0.08}$ and $\log(\kappa C_{1})\simeq -9.3$
(with a stringent 
upper limit $\simeq -3$, but unconstrained towards lower values)
where $\chi^2/{\rm d.f.}= 1.29$ (see left panel of Fig. 1).
Using Eq.(4) we can now relate the range of values of $\kappa C_{1}$
to the mass of the DM particle, from which we obtain that
\begin{equation}
M_x=\frac{1.205 \times 10^{-12}}{\kappa C_{1}} \frac{\langle \sigma u
  \rangle}{10^{-22}} \; {\rm GeV} \;,
\end{equation}
(see also right panel of Fig. 1) 
and since $\kappa C_{1}$ is unbound at small values, it is
consistent with currently 
accepted lower bounds of $M_x (\sim 10 {\rm GeV})$
(see \cite{cireli09} and references therein). 
The corresponding Hubble relation (Fig. 2), provided by the best-fit
model free parameters, is indistinguishable from that of the traditional
$\Lambda$CDM model, because of the very small value of $\kappa C_{1}
\simeq 10^{-9.3}$.
For completeness, we also show, as the dashed line, the IDM solution 
provided by $M_x \sim 1$eV ($\kappa C_{1} \simeq 10^{-3}$), 
which is the stringent lower bound found
by our analysis. In this case, the predicted Hubble expansion
deviates significantly from the traditional $\Lambda$ model 
at small $\alpha$ values indicating that it would probably create
significant alterations to the standard BBN 
(see \cite{Iocco09} and references therein).

\begin{figure}
  \includegraphics[height=.4\textheight]{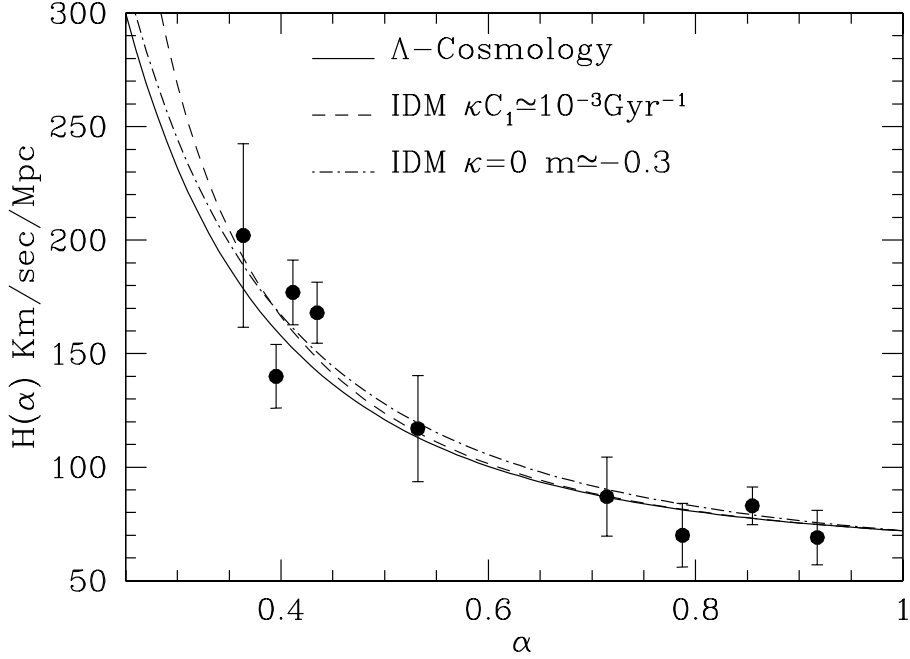}
\caption{Comparison of the Hubble function provided by the traditional
  $\Lambda$CDM model, which coincides with our $m=0$ model 
(for the best-fit model of the two free parameters - see text). 
The dashed line corresponds to our $m=0$ IDM model for the highest
$\kappa C_{1}$ bound,
provided by our fitting procedure ($\sim 10^{-3}$).
The dot-dashed line corresponds to our $\kappa=0$ IDM model 
for the best-fit model parameters ($m\simeq -0.3$ and
$\Omega_{2,0}\simeq 0.28$).
Finally, the points correspond to the observational data 
(\cite {simon05}).}
\end{figure}

Although the present analysis does not provide 
any important constraints on $M_x$
(within our model), we plan on the future to use a 
large amount of cosmologically relevant
data to attempt to place stronger $M_x$ constraints.

\section{Mimicking the Quintessence Dark Energy}
In this case, we assume that the annihilation term is 
negligible [$\kappa=0$ and $f(\alpha)$=0 and $F(\alpha)=1$] and the particle creation
term dominates. This situation is mathematically equivalent to the
gravitational DM particle creation process within 
the context of non-equilibrium thermodynamics  
\cite{Prigo1989},
the important cosmological consequences of which were studied by 
\cite{Lima2008} (see also references therein).

Using our nomenclature it is straightforward
to obtain the evolution of the total energy density from Eq.(\ref{sol1}), as:
\begin{equation}
\rho(\alpha)=D\alpha^{-3}+C_{1} \alpha^{m} \;,
      \label{sol77} 
\end{equation} 
where $D=1/C_{2}$.
The conditions in 
which the current model acts as a quintessence cosmology, are given by
$C_{1}>0$, $C_{2}>0$, and 
$w_{\rm IDM}=-1-m/3$, which implies that to have an inflection point,
the following should be satisfied: $w_{\rm IDM}<-1/3$ or $m>-2$ (see
section 2.1).
We note, that the Hubble flow is now given by
\begin{equation}
\left(\frac{H}{H_0}\right)^2 = \Omega_{\rm 2,0} \alpha^{-3}+ 
\Omega_{1,0} \alpha^{m} \;,
\end{equation}
where $\Omega_{\rm 2,0}= 8 \pi G D/3H_0^2$ and $\Omega_{1,0}=8
\pi G C_{1}/ 3H^2_0$.
Finally, by minimizing the corresponding $\chi^{2}$, 
we find that the best-fit model values are $\Omega_{\rm 2, 0}\simeq 0.28$
and $m\simeq -0.30$ ($w_{\rm IDM}\simeq -0.90$) with $\chi^2/{\rm d.f.}= 1.29$. 
The corresponding Hubble flow curve is shown in Fig. 2 as
the dot-dashed line. We note that
this solution is mathematically equivalent to that of the gravitational matter
creation model of \cite{Lima2008}.

\section{Conclusions}
In this work we investigate analytically the evolution of the global 
density of the universe
in the context of an interacting DM scenario by solving
analytically the collisional Boltzmann equation in an expanding
Universe. The possible disequilibrium between the DM particle creation and 
annihilation processes, regardless of its cause and in which 
the particle creation term dominates,
creates an effective source
term with negative pressure which (acting as {\em dark energy}) provides
an accelerated expansion phase of the scale factor.
Finally, comparing the observed Hubble function 
of a few high-redshift elliptical
galaxies with that predicted by our simplest IDM 
model ($m=0$), we find that the effective
annihilation term is quite small.

%\appendix

\section{Appendix A: The main assumptions of our model}
The reason why a cosmological constant or a component mimicking it
leads to a cosmic acceleration is because it introduces in 
Friedmann's equation, which governs the global
dynamics, a component which has an equation of state with negative
pressure. Our model
creates exactly the equivalent to the above, ie., an effective source
term which has negative pressure.
The avenue throw which this is accomplished is via the collisional
Boltzmann equation in an expanding universe (eg. \cite{Kolb1990})
in which the disequilibrium between the annihilation ($\kappa \rho^2$)
and particle creation ($\Psi$) processes provides the effective pressure: 
$P=(\kappa \rho^2 - \Psi)/3H$, which when negative acts exactly as a
repulsive force, and therefore provides a cosmic acceleration.
The basic assumption of this work is that the DM fluid is non-perfect,
ie., there is a disequilibrium between the particle pair annihilation and creation
processes. Note that in our general solution (eq.\ref{sol1}) we do
not specify the direction of this disequilibrium. Only in the
detailed analysis we seek solutions which provide an accelerated
phase of the universe (for which of course one condition is $P<0$),
which we indeed find.

\section{Appendix B: Solutions of the Riccati equation}
With the aid of the differential equation theory we present
solutions that are relevant to our eq.(\ref{bol2}). In general a Riccati 
differential equation is given by 
\begin{equation}
y^{'}=f(x)y^{2}+g(x)y+R(x)
\end{equation}
and it is fully solvable only when a particular solution is known.
Below we present two case in which analytical solutions are possible.

\begin{itemize}
\item {\it Case 1:} For the case where:

\begin{equation}
R(x)=C_{1}nx^{n-1}-C_{1}^{2}x^{2n}f(x)-C_{1}x^{n}g(x)
\end{equation}
the particular solution is $x^{n}$ and thus the corresponding
general solution can be written as:
\begin{equation}
y(x)=C_{1}x^{n}+\Phi(x)\left[C_{2}-\int_{1}^{x} f(u)\Phi(u)du \right]^{-1}
\end{equation}
where
\begin{equation}
\Phi(x)={\rm exp}\left[\int_{1}^{x} \left(2 C_{1}u^{n}f(u)+g(u)\right)du \right]
\end{equation}
and $C_{1}, C_{2}$ are the integration constants.
Using now eq. (\ref{fun1}) we get 
$\Psi(x)=xH(x)R(x)=C_{1}(n+3)x^{n} H(x)+\kappa C_{1}^{2}x^{2n}$.

\item {\it Case 2:} For the case where:
\begin{equation}
R(x)=h^{'}(x) \;\;\;\; {\rm with}\;\;\;\;g(x)=-f(x)h(x)
\end{equation}
the particular solution is $h(x)$ [in our case we have $h(x)=-3\kappa^{-1}H(x)$].  
The general solution now becomes:
\begin{equation}
y(x)=h(x)+\Phi(x)\left[C_{2}-\int_{1}^{x} f(u)\Phi(u)du \right]^{-1}
\end{equation}
where
\begin{equation}
\Phi(x)={\rm exp}\left[\int_{1}^{x} f(u)h(u)du\right] \;\;.
\end{equation}
In this framework, using eq. (\ref{fun1}) we finally get 
$\Psi(x)=xH(x)R(x)=-3\kappa^{-1}xH(x)H^{'}(x)$.
\end{itemize}
Note that the solution of {\it Case 1} 
is the only one providing a $\propto \alpha^{-3}$ 
dependence of the scale factor 
(see eqs. \ref{sol4a}, \ref{eq:lam} and \ref{sol77}).

\section{Appendix C: Justification of the functional form of  $\Psi$}
We assume that we have a non-perfect cosmic fluid in a  
disequilibrium phase with energy density
$\rho$ then from the collisional Boltzmann equation,
we have that
\begin{equation}
\Psi=\dot{\rho}+3H \rho+\kappa \rho^{2}=
\frac{{\rm d} \rho}{{\rm d}a} aH+3H \rho+\kappa \rho^{2}\;\;.
\end{equation}
Furthermore, we assume that for a convenient period of time,
the cosmic fluid, in an expanding Universe, is slowly diluted according to 
$\rho \sim C_{1}\alpha^{m}$ ($m\le 0$). From a mathematical point of view, 
the latter assumption simply means 
that a solution of the form $\propto \alpha^{m}$ is a particular solution 
of the Boltzmann equation. Therefore, we have finally that:
\begin{equation}
\Psi \simeq C_{1}(m+3)a^{m}H+\kappa C^{2}_{1}a^{2m} \;\;.
      \label{bol7} 
\end{equation}

\end{document}